\documentclass[preprint]{ptephy_v1}

\preprintnumber{OU-HET-1027}

\usepackage{bm}

\begin{document}

\title{Atiyah-Patodi-Singer index on a lattice}


\author{Hidenori Fukaya}
\author{Naoki Kawai}
\author{Yoshiyuki Matsuki}
\author{Makito Mori}
\affil{Department of Physics, Osaka University, Toyonaka 560-0043, Japan}
\author[1,2]{Katsumasa Nakayama}
\affil{NIC, DESY Zeuthen, Platanenallee 6, 15738 Zeuthen, Germany}
\author[1]{Tetsuya Onogi}
\author[1]{Satoshi Yamaguchi}


\begin{abstract}%
  We propose  a nonperturbative formulation of the Atiyah--Patodi--Singer(APS) index
  in lattice gauge theory in four dimensions, in which the index is given by 
  the $\eta$ invariant of
  the domain-wall Dirac operator. 
  Our definition of the index is always an integer with a finite lattice spacing.
  To verify this proposal, using
  the eigenmode set of the free domain-wall fermion,
  we perturbatively show in the continuum limit
  that the curvature term in the APS theorem appears
  as the contribution from the massive bulk
  extended modes, while the boundary $\eta$ invariant 
  comes entirely from the massless edge-localized modes.
\end{abstract}


\maketitle

\section{Introduction}
\label{sec:intro}

The topology of gauge fields, which plays a special role
in particle physics, is not well defined
in lattice gauge theory, where the space-time is
discretized and the notion of the ``manifold'' is lost.
However, from the viewpoint of fiber bundles,
the fiber space or the space of gauge fields
is still continuous and some geometrical properties remain.
For example, imposing a smoothness condition
on the link variables, 
the topological charge can be defined on a lattice \cite{Luscher:1981zq}.

A breakthrough was made by Hasenfratz {\it et al}. \cite{Hasenfratz:1998ri}
where they formulated the Atiyah--Singer(AS) index theorem \cite{Atiyah:1968mp}
on a finite lattice,
using a lattice Dirac operator
obeying the Ginsparg--Wilson(GW) relation \cite{Ginsparg:1981bj}.
The most popular choice is the overlap Dirac operator \cite{Neuberger:1997fp} 
\begin{eqnarray}
aD_\mathrm{ov} \equiv  1 + \gamma_5 \frac{H_\mathrm{W}}{\sqrt{H_\mathrm{W}^2}},  \;\;\; H_\mathrm{W} = \gamma_5(D_\mathrm{W}-m_0),
\label{eq:Dov}
\end{eqnarray}
where the lattice spacing is denoted by $a$,
$D_\mathrm{W}$ is the Wilson--Dirac operator, $\gamma_5$ is the chirality operator, and $m_0$ is the cutoff scale mass,
often chosen to be $1/a$.
The index is defined as
\begin{eqnarray}
  \label{eq:ASindex}
I
={\rm Tr}\gamma_5(1-a D_\mathrm{ov}/2).
\end{eqnarray}
The chiral symmetry \cite{Luscher:1998pqa}
through the GW relation 
$
D_\mathrm{ov}\gamma_5 + \gamma_5 D_\mathrm{ov} = a D_\mathrm{ov}\gamma_5 D_\mathrm{ov},
$
assures
that the contribution from the nonzero complex modes
are all cancelled, as are those from the doubler modes with the real eigenvalue $2/a$.
Thus $I$ is described by the zero modes only and
its equivalence to the AS index in the continuum limit
was confirmed in Refs. \cite{Kikukawa:1998pd, Luscher:1998kn, Fujikawa:1998if, Suzuki:1998yz,Adams:1998eg}.

In this paper we consider an extension of the AS theorem to manifolds with boundary,
known as the  Atiyah--Patodi--Singer(APS) index theorem \cite{Atiyah:1975jf},
\begin{eqnarray}
  Ind(D_{\rm APS})= 
  \frac{1}{32\pi^2} \int_{x_4>0} d^4x \epsilon^{\mu\nu\rho\sigma} {\rm tr} F_{\mu\nu}F_{\rho\sigma}-\frac{1}{2}\eta(iD^{3\mathrm{D}}),
\end{eqnarray}
where $F_{\mu\nu}$ is the curvature of gauge fields of $SU(N)$,
and the second term corresponds to the $\eta$ invariant
of the massless three-dimensional Dirac operator $iD^{3\mathrm{D}}$ on the boundary set at $x_4=0$, here.
The APS theorem is equally important in mathematics but its lattice version is not known yet.
This is because the APS index in the original setup
requires a nonlocal boundary condition, which is difficult to put on a lattice.
In fact, any other physically sensible boundary condition is not compatible with the chiral symmetry \cite{Luscher:2006df}.
Besides, there has been no strong motivation to consider
the APS index in physics, as we have not been that interested in a
space-time with boundaries.

Recently, the APS index theorem has been drawing attention from physics,
as it was pointed out that the theorem describes the
bulk--edge correspondence of topological insulators \cite{Witten:2015aba},
through the cancellation of the time-reversal symmetry anomaly.
Motivated by this fact, three of the authors proposed
a new formulation \cite{Fukaya:2017tsq} of the APS index
in four-dimensional continuum theory.
Three of us used the domain-wall fermion formulation \cite{Jackiw:1975fn, Callan:1984sa, Kaplan:1992bt},
which offers a very similar setup to the topological insulators.
This new formulation is quite different from the original definition by APS, but
the equivalence of the new formulation with the original APS was 
mathematically proved in Ref. \cite{Fukaya:2019qlf} in continuum theory.
In this work we try to apply this new formalism to the lattice gauge theory.

The rest of the paper is organized as follows.
In Sect.~\ref{sec:eta} we discuss the new formulation of the index
in continuum theory \cite{Fukaya:2017tsq, Fukaya:2019qlf} and show that
the lattice formulation is quite straightforward.
The key to obtaining the continuum limit of the lattice APS index
is to construct the free domain-wall complete set, shown in Sect.~\ref{sec:completeset}.
Then we separately evaluate the bulk and edge contributions
to the total index in Sect.~\ref{sec:evaluation}.
We give a conclusion in Sect.~\ref{sec:conclusion}.

\section{Rewriting the index by the $\eta$ invariant of the massive Dirac operator}
\label{sec:eta}

Let us go back to the AS index on a four-dimensional periodic lattice. 
Simply substituting Eq.~(\ref{eq:Dov}) into Eq.~(\ref{eq:ASindex}),
one notices that the index is
equivalent to the quantity
\begin{equation}
  \label{eq:ASeta}
I=
 -\frac{1}{2}\eta(H_\mathrm{W}) \equiv -\frac{1}{2}{\rm Tr}\frac{H_\mathrm{W}}{\sqrt{H_\mathrm{W}^2}}.
\end{equation}
In fact, Eq.(\ref{eq:ASeta}) gives a good definition of the $\eta$ invariant of $H_\mathrm{W}$,
which is a regularized number of positive eigenmodes minus the number of negative eigenmodes.
This fact is suggestive, in that the index may be defined by the massive
Wilson--Dirac operator, without chiral symmetry or the GW relation at all.
In fact, such attempts \cite{Itoh:1987iy} were made in studies before formulation of the overlap fermions.
However, little was discussed about the possibility that the $\eta$ invariant of the massive Dirac operator is as equally important a quantity as the original index in mathematics.

Reference \cite{Fukaya:2019qlf} showed that in continuum theory, for any APS index
on an even-dimensional Riemannian manifold with boundary,
there exists a domain-wall operator
acting on a closed manifold without boundary,
sharing its ``half'' with the original setup with APS,
and its $\eta$ invariant is equal to the original APS index.\footnote{
Our domain-wall fermion here is in four dimensions,
which is one dimension lower than the standard five-dimensional ones.
As the chirality operator exists only in even dimensions,
the mathematical consequences are quite different from those in
five-dimensions. For example, in Ref.\cite{Kaplan:1995pe},
the $\eta$ invariant appears as a non-integer quantity
to describe the phase of the Weyl fermion at the edge,
while our target is Dirac fermion whose determinant is real.
  }
For example, let us consider the four-dimensional Dirac operator
$H^c_{\mathrm{DW}}=\gamma_5(D-M{\rm sgn}(x_4))$ where the mass term changes its sign
at the boundary $x_4=0$.
Then, we can prove with the Pauli--Villars regularization that
\begin{eqnarray}
  -\frac{1}{2}\eta(H^c_{DW})^{PVreg.}= Ind(D_{\rm APS}).
\end{eqnarray}
It is interesting to note that the new formulation
is given on a closed manifold, but the result reflects only ``half'' of it in the region $x_4>0$.\footnote{
When we flip the sign of the Pauli-Villars mass, we obtain the APS index in the region $x_4<0$.
}
Note that the mathematical setups between the domain-wall fermion and original APS
are totally different: the original APS boundary condition completely kills the positive modes of
the boundary Dirac operator, while the domain-wall fermion can
live even ``outside'' of the boundary, or the region in $x_4<0$.
The APS boundary condition never allows edge-localized modes to exist,
while the boundary $\eta$ invariant of the domain-wall system
comes entirely from the edge modes of the domain-wall.

The proof includes the case with no domain-wall or boundary:
the AS index is 
\begin{equation}
  I = -\frac{1}{2}\eta(\gamma_5(D-M))^{\mathrm{PVreg.}},
\end{equation}
where the Pauli--Villars mass has the opposite sign to $M$.
Therefore, the $\eta$ invariant of the massive Dirac operator
is an equally (or possibly more) fundamental quantity than the original index
of the ``massless'' Dirac operator.
We should note that the chiral symmetry is lost in
the massive Dirac operator, and yet the index can be defined.
The lattice index theorem described in Sect.~\ref{sec:intro} ``knew'' this fact,
and the Wilson--Dirac operator can
define the AS index on a lattice by $-\eta(H_\mathrm{W})/2$, which
happens to be the same as that given by the massless $D_{\mathrm{ov}}$.

It is now natural to speculate that the
Wilson--Dirac operator with a mass term having kink structures,
\begin{eqnarray}
  H_{\mathrm{DW}} = \gamma_5 \left[D_\mathrm{W}-M_1\epsilon\left(x_4+\frac{a}{2}\right)\epsilon\left(L_4-\frac{a}{2}-x_4\right)+M_2\right],
\end{eqnarray}
where $\epsilon(x)={\rm sgn}(x)$, 
may have an interesting relation to topology.
Indeed, the goal of our paper is to
show that the $\eta$ invariant of $H_{DW}$
gives a nonperturbative formulation of the APS index theorem on a lattice.
Here, we assume a periodic boundary 
identifying $x_4=L_4$ and $x_4=-L_4$, setting $L_4/a$ an integer,
giving two domain-walls at $x_4=-a/2$ and $x_4=L_4-a/2$.
We assume $M_1>M_2>0$.
For the other three directions we use a periodic boundary condition with the same periodicity $L$.

It is good to know that $\eta(H_{\mathrm{DW}})$ is
guaranteed to be
 an integer 
by definition,
 since $H_{\mathrm{DW}}$ is
a Hermitian operator on  
 a finite-dimensional
  vector space.
Moreover, under the condition that no eigenvalue of $H_{\mathrm{DW}}$ crosses zero, 
its variation is always zero,
\begin{eqnarray}
  \label{eq:DeltaHDW}
  \delta \eta(H_{\mathrm{DW}}) = 0.
\end{eqnarray}

In the rest of this paper, we will perturbatively show that
\begin{eqnarray}
  \label{eq:latAPSgeom}
  -\frac{1}{2}\eta(H_{\mathrm{DW}}) =& \displaystyle{\frac{1}{32\pi^2} \int_{0<x_4<L_4} d^4x \epsilon^{\mu\nu\rho\sigma} {\rm tr} F_{\mu\nu}F_{\rho\sigma}}
  -\frac{1}{2}\eta(iD^{3\mathrm{D}})|_{x_4=0}+\frac{1}{2}\eta(iD^{3\mathrm{D}})|_{x_4=L_4},
\end{eqnarray}
up to $O(a)$ corrections.
To simplify the computation, we make the coefficient of the Wilson term unity,
and consider the $M_1+M_2\to \infty$ limit with $M_1-M_2=M$ fixed.
This case corresponds to the Shamir domain-wall \cite{Shamir:1993zy,Furman:1994ky}.
In the continuum theory \cite{Fukaya:2017tsq},
the result is shown to be independent of $M_2$.
Then we take the hierarchical scaling limit $|\lambda_{\rm edge}|\ll M\ll 1/a$, where
$\lambda_{\rm edge}$ denotes a typical eigenvalue of  low-lying edge-localized modes.
In this limit, the bulk modes have large energy and their correlations
exponentially decay in every direction.
Therefore, the density of the $\eta$ invariant near $x_4=0$ 
can be locally evaluated using a complete set of semi-infinite space-time in $x_4\ge 0$,
and simply interpolated to the result obtained near $x_4=L_4$. 
For the same reason, we also treat the momenta in other directions as  continuous.

\section{Free domain-wall fermion complete set}
\label{sec:completeset}

Let us consider the 
eigenproblem
 of $a^2H_{\mathrm{DW}}^2$
for the free fermion, taking the $L_4=\infty$ limit.
We can take the direct product of the solutions in
the $x_{i=1,2,3}$ directions as the plane wave
 $\psi^{\rm 3D}_{p}(\bm{x}) = e^{i\bm{p}\cdot \bm{x}}/\sqrt{(2\pi)^3}$,
having two-spinor components,
and that in the $x_4$ direction.
We denote the three-momentum by $\bm{p}=(p_1,p_2,p_3)$.
Using the abbreviations $s_i=\sin(p_i a)$ and $c_i=\cos(p_i a)$,
the squared domain-wall Dirac operator is expressed by
\begin{eqnarray}
  a^2 (H_{\mathrm{DW}}^{\rm 0})^2 &=& s_i^2+
  \theta(x_4+a/2)\{M_+^2-(1+M_+)(a^2 \nabla^*_4\nabla_4)\}\nonumber\\
  && + \theta(-x_4-a/2)\{M_-^2-(1+M_-)(a^2 \nabla^*_4\nabla_4)\}\nonumber\\&&
  + 2M_1a (P_+ \delta_{x_4, -a}S_4^+ - P_-\delta_{x_4, 0}S_4^- ),\\
  M_\pm &=& \sum_{i=1,2,3} (1-c_i) \mp M_1a +M_2a,
\end{eqnarray}
where $\theta(x)=(\epsilon(x)+1)/2$ is the step function,
$\nabla_\mu$ and $\nabla_\mu^*$ denote the forward and backward
difference operators, respectively,  $P_\pm=(1+\gamma_4)/2$,
and $S_\mu^\pm$ is a shift operator
by the unit lattice vector $\hat{\mu}a$:
$S_\mu^\pm f(x)=f(x \pm a \hat{\mu})$.
We have three types of eigenfunctions (in the $x_4$ direction) of $a^2H_{\mathrm{DW}}^2$:
(i) edge-localized modes at $x_4=0$, (ii) extended modes but only for $x_4\ge 0$,
and (iii) extended modes at any $x_4$.

In the $M_1+M_2\to \infty$ limit with $M_1-M_2=M$ fixed,
only 
the eigenmodes of types
(i) and (ii) survive to form a complete set.
They coincide with those for the Shamir domain-wall fermion 
\cite{Shamir:1993zy, Furman:1994ky}
on which the simple Dirichlet boundary condition
$\phi(x_4) = 0$ for $x_4<0$ is imposed.
More explicitly, we have
\begin{eqnarray}
  \phi_-^{\rm edge}(x_4) &=& \sqrt{-M_+(2+M_+)/a}e^{-K x_4},\\
  \label{eq:edge}
\phi_+^\omega(x_4) &=& \frac{1}{\sqrt{2\pi}}(e^{i\omega (x_4+a)}-e^{-i\omega (x_4+a)}),\\
\phi_-^\omega(x_4) &=& \frac{1}{\sqrt{2\pi}}(C_\omega e^{i\omega x_4}-C_\omega^* e^{-i\omega x_4})
\label{eq:bulk}
\end{eqnarray}
in the region $x_4\geq 0$, where the subscript $\pm$ denotes the eigenvalue of $\gamma_4=\pm 1$,
\begin{eqnarray}
  K &=& - \ln (1+M_+)/a,\\
  C_\omega &=& - \frac{(1+M_+)e^{i\omega a}-1}{|(1+M_+)e^{i\omega a}-1|},
\end{eqnarray}
and the eigenvalue of $H_{\mathrm{DW}}^2$ is
\begin{eqnarray}
  \Lambda^2 = \left\{
  \begin{array}{cc}
    s_i^2/a^2 & (\mbox{edge})\\
    (s_i^2 +M_+^2-2(1+M_+)(\cos \omega a -1))/a^2 & (\mbox{bulk})
  \end{array}
  \right. 
\end{eqnarray}
Here 
we choose the fermion mass
in the range $0<Ma<2$, which is required
from the normalizability of the edge-localized modes,
or $|1+M_+|<1$.
This condition eliminates the contribution from the doubler modes,
which have $|\bm{p}|a >\pi$.
The edge modes can exist only in the $\gamma_4=-1$ sector.
With this complete set, 
we can
separately evaluate the contributions from the bulk and edge modes.

One can easily confirm that the eigenmode set above satisfy
the orthonormal condition,
\begin{eqnarray}
  a\sum_{x_4=0}^\infty \phi_-^{\rm edge}(x_4)^\dagger \phi_-^{\rm edge}(x_4) &=& 1,\\
  a\sum_{x_4=0}^\infty \phi_\pm^{\omega'}(x_4)^\dagger \phi_\pm^{\omega}(x_4) &=& \delta(\omega-\omega'),
\end{eqnarray}
where the summation over $x_4$ is taken for integer multiples of $a$,
and the completeness,
\begin{eqnarray}
  \sum_{g=\pm}\int_0^{\pi/a}d\omega u_g\phi_g^\omega (x_4)\phi_g^\omega (x_4')^\dagger u_g^\dagger
  + u_-\phi_-^{\rm edge}(x_4)\phi_-^{\rm edge}(x_4')^\dagger u_-^\dagger 
  = \frac{\delta_{x_4,x_4'}}{a} I_{2\times 2},
\end{eqnarray}
where $u_\pm$ and $I_{2\times 2}$ are the eigenvectors
and the $2\times 2$ identity matrix in the eigenvector space of $\gamma_4$, respectively.

\section{Evaluation of the $\eta$ invariant of the domain-wall operator}
\label{sec:evaluation}

In this section we perturbatively evaluate the $\eta$ invariant
of the domain-wall Dirac operator, 
and show its  convergence to the APS index.
We separately evaluate the bulk and edge contributions
and confirm their parity (or T) anomaly cancellation.

For the bulk contribution, we consider the 
density of the $\eta$ invariant:
\begin{eqnarray}
  -\frac{1}{2}{\rm tr} \frac{H_\mathrm{DW}}{\sqrt{H_\mathrm{DW}^2}}(x)^{\rm bulk} 
 &=&
 -\left.
 \frac{1}{2}
  \sum_{g=\pm} \int_0^{\pi/a} d\omega \int_{-\pi/a}^{\pi/a}d^3p
  \right\{\nonumber\\&&\left.
   [\psi^{\rm 3D}_{p}(\bm{x})\otimes \phi^\omega_g(x_4)]^\dagger
  {\rm tr}\left[P_g\frac{H_\mathrm{DW}}{\sqrt{H_\mathrm{DW}^2}}P_g\right][\psi^{\rm 3D}_{p}(\bm{x})\otimes\phi^\omega_g(x_4)]\right\}, \nonumber\\
\end{eqnarray}
rather than its integral form. 
Here, the trace ${\rm tr}$ is taken over color and spinor indices only.
Since every bulk mode has a larger energy than $M_+^2$,
the density is expressed as a local function.
The analysis below is similar to that for the periodic lattice studied in Ref. \cite{Suzuki:1998yz}.
The dependence on the gauge link variables is perturbatively treated as
\begin{eqnarray}
  H_{DW}^2 &=& (H_\mathrm{DW}^{\rm 0})^2 + \Delta H_\mathrm{DW}^2,\\
  \Delta H_\mathrm{DW}^2 &=& -\frac{1}{4}\sum_{\mu,\nu}[\gamma^\mu,\gamma^\nu][\tilde{D}_\mu, \tilde{D}_\nu] - \gamma^\mu[\tilde{D}_\mu,\tilde{R}],  
\end{eqnarray}
where we have neglected terms having no $\gamma_\mu$'s, which 
do not contribute to the index,
and we can express 
\begin{eqnarray}
  \tilde{D}_\mu &=& \frac{1}{2a}\left[e^{ip_\mu a}(U_\mu(x) S^+_\mu-1) -e^{-ip_\mu a}(S^-_\mu U_\mu(x)^\dagger -1)\right],
  \\
  \tilde{R} &=& -\frac{1}{2a}\sum_\mu \left[e^{ip_\mu a}(U_\mu(x) S^+_\mu-1)\right. 
    \left.
    + e^{-ip_\mu a}(S^-_\mu U_\mu(x)^\dagger -1)\right],
\end{eqnarray}
assuming they operate on $[\psi^{\rm 3D}_{p}(\bm{x})\otimes\phi^{\omega}_\pm(x_4)]$.

Note that $(H_\mathrm{DW}^{\rm 0})^2$ and $\Delta H_\mathrm{DW}^2$ do not commute but
their commutator either increases the order of $a$ (by derivatives)
or is localized at $x_4=0, a$ (by $\gamma_4$), to which
the bulk mode's propagation is exponentially suppressed. 
Therefore, we can expand $1/\sqrt{H_\mathrm{DW}^2}$ as if the operators were all commuting.
Noting the existence of $\gamma_4$ in the projection $P_\pm$, and that of $\gamma_5$ in the numerator,
those terms having three or four gamma matrices can survive the spinor trace.
After some lengthy but straightforward computations 
we find that only the first two terms in the following expression are relevant,
\begin{eqnarray}
  \frac{H_\mathrm{DW}}{\sqrt{H_\mathrm{DW}^2}} &=&
  \frac{3a^4}{8}{\rm tr}\left[\gamma_5M_+' \frac{\gamma_\mu\gamma_\nu\gamma_\rho\gamma_\sigma}{4(\Lambda^2 a^2)^{5/2}}
    [\tilde{D}^{\mu},\tilde{D}^{\nu}][\tilde{D}^{\rho},\tilde{D}^{\sigma}](x)\right]
  \nonumber\\&&+ \frac{3a^4}{8}{\rm tr}\left[\gamma_5\frac{i\gamma_\mu s^\mu \gamma_\nu\gamma_\rho}{(\Lambda^2 a^2)^{5/2}}
    [\tilde{D}^{\nu},\tilde{D}^{\rho}][\gamma_\sigma \tilde{D}^{\sigma},\tilde{R}](x)\right]
  \nonumber\\&&+\cdots,
\end{eqnarray}
where $M_+' = M_++(1-c_4)$ with $c_4=\cos \omega a$, and all Greek indices are summed.

Substituting the explicit forms of the free domain-wall fermion complete set, and using the standard expansion
$U_\mu(x) = \exp(ia A_\mu)$,
\begin{eqnarray}
  [\tilde{D}^{\mu},\tilde{D}^{\nu}] &=& i  c_\mu c_\nu F_{\mu\nu} + O(a),\\
 {}[\tilde{D}^{\mu},\tilde{R}] &=& i  c_\mu \sum_\nu s_\nu F_{\mu\nu} + O(a),
\end{eqnarray}
we obtain 
\begin{eqnarray}
  -\frac{1}{2}{\rm tr} \frac{H_\mathrm{DW}}{\sqrt{H_\mathrm{DW}^2}}(x)^{\rm bulk} =
  (I(M)+I^\mathrm{DW}(M,x_4))\frac{1}{32\pi^2} \epsilon^{\mu\nu\rho\sigma} {\rm tr} F_{\mu\nu}F_{\rho\sigma}(x),
\end{eqnarray}
up to $O(a)$ corrections, where
\begin{eqnarray}
  I(M) &=& \frac{3a^4}{8\pi^2}\int_{-\pi/a}^{\pi/a}d\omega d^3 p
  \prod_\mu c_\mu \frac{(-M_+'+\sum_\nu s_\nu^2/c_\nu)}{(\Lambda^2 a^2)^{5/2}},
\end{eqnarray}
which was already evaluated in Ref. \cite{Suzuki:1998yz} and $I(M)=1$
in the continuum limit, for our choice $0<Ma<2$.
The $x_4$-dependent part, 
\begin{eqnarray}
  \label{eq:IDW}
  I^\mathrm{DW}(M,x_4) &=& \frac{3a^4}{8\pi^2}\int_{-\pi/a}^{\pi/a}d\omega d^3 p
  \prod_\mu c_\mu \frac{(-M_+'+\sum_\nu s_\nu^2/c_\nu)}{(\Lambda^2 a^2)^{5/2}}
  \left(-\frac{C_\omega^2+e^{2i\omega a}}{2}e^{2i\omega x_4}\right),
\end{eqnarray}
is a special contribution due to the domain-wall, 
but we can show that its integral with the curvature contribution
over $x_4$ in the continuum limit is suppressed by a factor of $1/M$ as
\begin{eqnarray}
  \label{eq:IDWbound}
  \left|\int dx_4 I^\mathrm{DW}(M,x_4)\epsilon^{\mu\nu\rho\sigma} {\rm tr} F_{\mu\nu}F_{\rho\sigma}(x) \right|
  < \frac{3\left|\epsilon^{\mu\nu\rho\sigma} {\rm tr} F_{\mu\nu}F_{\rho\sigma}\right|^{\max}(\bm{x})}{8M},
\end{eqnarray}
where $|O|^{\max}(\bm{x})$ is the absolute maximum of
the function $O(x)$ along the string extending in the $x_4$ direction
located at $\bm{x}=(x_1,x_2,x_3)$.
See Appendix~\ref{app:IDW} for the details.
Therefore, we can conclude $I(M)+I^\mathrm{DW}(M,x_4)=1+O(1/M)$
and the bulk contribution is the standard curvature term.

Next, let us evaluate the edge-localized contribution.
To this end, we reconsider the eigenproblem for the edge modes
with nontrivial gauge link variables, assuming
its mild $x_4$ dependence compared to $1/M$.
More explicitly, we have, in the $U_4=1$ gauge,
\begin{eqnarray}
  H_\mathrm{DW}\phi(x)&=&\Lambda\phi(x),
\end{eqnarray}
\begin{eqnarray}
  \label{eq:HDWx4dep}
  H_\mathrm{DW} &=&\gamma_5\left[-P_-\nabla_4+P_+\nabla_4^* +\gamma_iD^i(x_4)+M_+(x)/a)\right],\\
  M_+(x)/a &=& -\frac{1}{2a}\sum_{i=1,2,3} \left[(U_i(x) S^+_i-1)\right. 
    \left.
    + (S^-_i U_i(x)^\dagger -1)\right]-M,
\end{eqnarray}
where $D^i(x_4)$ is the symmetrized spatial covariant difference operator at a slice $x_4$.
Note that $D^i(x_4)$ depends on $x_4$ through the link variables.
$M_+(x)$ is also $x_\mu$ dependent through the link variables.
In the following, we assume that the eigenvalue $\Lambda$ is low
compared to $M$ and the mass of the doublers modes, where we can approximate $M_+(x)/a=-M+O(a)$
and ignore the position dependence.

At the leading order of the adiabatic approximation,
we have a solution of the form
\begin{equation}
  \phi(x) = \phi^{\rm 3D}_{\lambda(0)}(\bm{x}) \otimes \phi^\mathrm{edge}(x_4),
\end{equation}
where $\phi^{\rm 3D}_{\lambda(0)}(\bm{x})$ is an eigenstate of
$i\sigma_i D^i(x_4=0)$ ($\sigma_i$ are the Pauli matrices) with the eigenvalue $\lambda(0)$,
and 
\begin{eqnarray}
  \label{eq:edge'}
  \phi^\mathrm{edge}_-(x_4) = \sqrt{M(2-Ma)} e^{-K x_4},
\end{eqnarray}
where $e^{-K a}=(1-Ma)$ and the eigenvalue of $\gamma_4$ is $-1$.
Here the Dirac representation ($\gamma_4={\rm diag}(1,-1)$) is employed.
On the above solution, the domain-wall operator acts as
$H_\mathrm{DW}=\gamma_5\gamma_iD^i = {\rm diag}(0, i\sigma_iD^i(x_4=0))$.
Therefore, the eigenvalue $\Lambda$ essentially equals $\lambda(0)$.
Since $K=M+O(a)$, the orthonormality with the bulk modes which was
given in terms of free domain-wall fermion is guaranteed in the continuum limit.
In Appendix~\ref{app:orth}, we show the details,
explicitly evaluating the inner product with the free bulk fermion modes.

In fact, this leading-order solution is enough to evaluate the
edge mode part, as the exponential dumping of the eigenfunctions allows us
to expand the operator in $x_4$, and its dependence
is suppressed as
\begin{equation}
  \sum_{x_4} x_4^n e^{-2Kx_4} = \frac{1}{(-2)^n}\frac{d^n}{dK^n}\frac{1}{1-e^{-2Ka}}\sim  1/M^{n+1}.
\end{equation}
For example, in the $x_4$ expansion of $\gamma_iD^i(x_4)$
in Eq.~(\ref{eq:HDWx4dep}), the linear contribution in $x_4$
to the $\eta$ invariant ${\rm Tr}_\mathrm{edge}\frac{H_\mathrm{DW}}{\sqrt{H_\mathrm{DW}^2}}$
is suppressed as
\begin{eqnarray}
  a\sum_{x_4} \phi^\mathrm{edge}_-(x_4)^\dagger
  \frac{x_4 \partial_{x_4}\gamma_5\gamma_iD^i(x_4=0)}{\sqrt{\lambda(0)^2}}
  \phi^\mathrm{edge}_-(x_4)
  &=& \frac{a \lambda^\prime(0)}{\sqrt{\lambda(0)^2}}
  M(2-Ma) \sum_{x_4} x_4 e^{-2Kx_4}
  \nonumber\\
  &=&-\frac{\lambda^\prime(0)}{\sqrt{\lambda(0)^2}}
  \frac{Ma^2(2-Ma)(1-Ma)}{(1-(1-Ma)^2)^2}
  \nonumber\\
  &\sim& -\frac{\lambda^\prime(0)}{\sqrt{\lambda(0)^2}}
  \frac{1}{2M},
\end{eqnarray}
where we have taken the $a\to 0$ limit,
and $\lambda^\prime(0)$ is the $x_4$ derivative
of the eigenvalue at $x_4=0$ in the adiabatic evaluation.
Therefore, if we take $M$ to be big enough compared
to the derivative of the gauge fields,
the leading adiabatic evaluation is valid.

Now we are ready to compute the edge mode's contribution to the $\eta$ invariant.
We emphasize again that the eigenvalue $\Lambda$ of the domain-wall Dirac operator
coincides with $\lambda(0)$, the eigenvalue of the boundary Dirac operator.
Then the edge mode part becomes
\begin{eqnarray}
  -\frac{1}{2}{\rm Tr}_\mathrm{edge}\frac{H_\mathrm{DW}}{\sqrt{H_\mathrm{DW}^2}} &=&
  -\sum_{\lambda(0)} \frac{{\rm sgn}\lambda(0)}{2}
  = -\frac{1}{2}\eta(i\sigma_i D^i)|_{x_4=0}.
\end{eqnarray}
Note, however, that the above approximation does not hold for higher energy
$\lambda(0) \sim M$ where the bulk and edge modes mix.
Therefore, the edge mode part alone is not an integer, and it is difficult to
separate the edge and bulk contributions in such an energy region.

In order to evaluate the noninteger part of
the edge-localized contribution,
let us take the variation with respect to the link variables.
From Eq.~(\ref{eq:DeltaHDW}), we have the explicit ``bulk-edge correspondence''
\begin{eqnarray}
  \label{eq:bulkedge}
-\frac{1}{2}\delta {\rm Tr}_\mathrm{edge}\frac{H_\mathrm{DW}}{\sqrt{H_\mathrm{DW}^2}} &=& \frac{1}{2}\delta {\rm Tr}_\mathrm{bulk}\frac{H_\mathrm{DW}}{\sqrt{H_\mathrm{DW}^2}},
\end{eqnarray}
where ${\rm Tr}_\mathrm{bulk/edge}$ is the trace taken over the bulk/edge modes only.
Thanks to the locality of the gapped bulk modes, the right-hand side is much easier to
perturbatively compute, leading to 
\begin{eqnarray}
  = -\frac{1}{16\pi^2}\int d^3x \delta {\rm tr}_c
  \left[\varepsilon_{0\nu\rho\sigma}\left(A^\nu \partial^\rho A^\sigma +\frac{2i}{3}A^\nu A^\rho A^\sigma\right)\right]. 
\end{eqnarray}
Namely, the noninteger part of the edge-localized contribution is the Chern--Simons action,
except for some extra gauge-invariant and constant contributions.\footnote{
  In the appendix of \cite{Fukaya:2017tsq},  an explicit example of such a
  constant and gauge invariant contribution is shown.
}
Thus, we can clearly see in Eq.~(\ref{eq:bulkedge}) the cancellation of the
parity or T anomaly between the bulk and edge states.

So far, we have neglected the finiteness of the fourth direction $L_4$.
Since its effect in the bulk is suppressed by $e^{-ML_4}$,
in the $\lambda_\mathrm{edge}\ll M\ll 1/a$ limit, we can safely interpolate
our result to that with the anti-domain-wall at $x_4=L_4-a/2$,
and obtain the desired result already shown in Eq.~(\ref{eq:latAPSgeom}).

\section{Conclusion}
\label{sec:conclusion}

We have shown that the
$\eta$ invariant of the domain-wall fermion Dirac operator
converges to the Atiyah--Patodi--Singer index in the continuum limit.
Our definition of the index is always an integer at finite lattice spacings,
therefore this achieves a nonperturbative formulation of the
APS index theorem on the lattice.
  In order to guarantee that this index always matches that in continuum,
  we will need some ``admissibility'' condition on the link variables, as is the case in the
  AS index \cite{Luscher:1981zq}; this, however, is beyond the scope of this work.
We have also explicitly demonstrated the cancellation of the
parity anomaly (or time-reversal symmetry anomaly)
by the Chern--Simons action in the contributions to the total index.
This is an essential property of the bulk-edge correspondence \cite{Witten:2015aba, Hatsugai:1993ywa}.

Acknowledgments: We thank M. Furuta, S. Matsuo, and M. Yamashita
for their fruitful feedback and encouragement from mathematics.
We thank H. Suzuki for his instruction on the 
computation
of $I(M)$.
We also thank S. Aoki and Y. Kikukawa for useful discussions.
This work was supported in part by JSPS KAKENHI grant number JP15K05054, JP18H01216, JP18H04484, JP18J11457, JP18K03620,
and JP19J20559.
The authors thank the Yukawa Institute for Theoretical Physics at Kyoto University. Discussions during the YITP workshop YITP-T-19-01 on ``Frontiers in Lattice QCD and related topics" were useful to completing this work. T.O. would also like to thank YITP for their kind hospitality during his stay.




\appendix

\section{Contribution from $I^{DW}(M,x_4)$}
\label{app:IDW}

In this appendix we explicitly show Eq.~(\ref{eq:IDWbound}).
First we note the phase factor $e^{2i\omega x_4}$ in Eq.~(\ref{eq:IDW}),
to which the singularity due to doublers gives a contribution
suppressed as $e^{-x_4/a}$.
Therefore, we can take a naive continuum limit only taking
the physical poles into account approximating
$s_\mu \sim p_\mu a$, and $c_\mu\sim 1$.
In this limit, $I^\mathrm{DW}(M,x_4)$ becomes
\begin{eqnarray}
  I^\mathrm{DW}(M,x_4) &=& 3M
  \int_0^\infty p^2 dp \int_{-\infty}^\infty \frac{d\omega}{2\pi}
  \frac{1}{(p^2+M^2+\omega^2)^{5/2}}
  \left[\frac{2iM}{\omega-iM}e^{2i\omega x_4}\right]
  \nonumber\\
  &=& 2M \int_0^\infty dt \sqrt{t} \frac{\partial^2}{\partial t^2} J(t,M,x_4),
\end{eqnarray}
where 
\begin{eqnarray}
J(t,M,x_4) &=& \int_{-\infty}^\infty \frac{d\omega}{2\pi}
  \frac{1}{(t+M^2+\omega^2)^{1/2}}
  \left[\frac{2iM}{\omega-iM}e^{2i\omega x_4}\right].
\end{eqnarray}  

We can formally integrate $J(t,M,x_4)$, which consists of two terms:
\begin{eqnarray}
  J(t,M,x_4) &=& J_1(t,M,x_4)+J_2(t,M,x_4),
  \nonumber\\
  J_1(t,M,x_4) &=& -\frac{2M}{\pi}e^{-2Mx_4}\left[\frac{1}{2\sqrt{t}}
  \arccos\left(\frac{M}{\sqrt{M^2+t}}\right)\right],\\
  J_2(t,M,x_4) &=& -\frac{2M}{\pi}e^{-2Mx_4}\left[
    \int_0^{x_4} dx_4' e^{2Mx_4'}K_0(2\sqrt{M^2+t}x_4')
    \right],
\end{eqnarray}
where $K_\nu(x)$ denotes the modified Bessel function.
For $J_1(t,M,x_4)$ it is not difficult to compute
\begin{eqnarray}
  2M \int_0^\infty dt \sqrt{t} \frac{\partial^2}{\partial t^2} J_1(t,M,x_4)
  &=& -\frac{1}{4} e^{-2Mx_4}.
\end{eqnarray}
For $J_2(t,M,x_4)$, let us take a partial integration to obtain
\begin{eqnarray}
  2M \int_0^\infty dt \sqrt{t} \frac{\partial^2}{\partial t^2} J_2(t,M,x_4)
  &=& - M \int_0^\infty dt \frac{1}{\sqrt{t}} \frac{\partial}{\partial t} J_2(t,M,x_4)
  \nonumber\\
  &&\hspace{-1.5in} =- \frac{2M^2e^{-2Mx_4}}{\pi}
  \left[
    \int_0^{x_4} dx_4' x_4'e^{2Mx_4'} \int_0^\infty dt \frac{1}{\sqrt{t}}
    \frac{K_1(2\sqrt{M^2+t}x_4')}{\sqrt{M^2+t}}
    \right].
\end{eqnarray}
Combining these results, we obtain
\begin{eqnarray}
  I^\mathrm{DW}(M,x_4) &=&
  -e^{-2Mx_4}\left[\frac{1}{4}+\frac{2M^2}{\pi}
    \int_0^{x_4} dx_4' x_4'e^{2Mx_4'} \int_0^\infty dt 
    \frac{K_1(2\sqrt{M^2+t}x_4')}{\sqrt{t}\sqrt{M^2+t}}
    \right].
\end{eqnarray}

We are now ready to show Eq.~(\ref{eq:IDWbound}).
Noting the fact that $I^\mathrm{DW}(M,x_4)$ is negative at any $x_4$,
we have the inequality
\begin{eqnarray}
  \left|\int dx_4 I^{DW}(M,x_4)\epsilon^{\mu\nu\rho\sigma} {\rm tr} F_{\mu\nu}F_{\rho\sigma}(x) \right|
  <
  \left|\epsilon^{\mu\nu\rho\sigma} {\rm tr}
  F_{\mu\nu}F_{\rho\sigma}\right|^{\max}(\bm{x})
  \int_0^\infty dx_4 |I^{DW}(M,x_4)|,  
\end{eqnarray}
where $|O|^{\max}(\bm{x})$ is the absolute maximum of
the function $O(x)$ along the string extending in the $x_4$ direction
located at $\bm{x}=(x_1,x_2,x_3)$.
The $x_4$ integral is analytically computable as follows:
\begin{eqnarray}
  \int_0^\infty dx_4 |I^\mathrm{DW}(M,x_4)|
  &=& \frac{1}{8M}
  + \frac{M}{\pi}
  \int_0^{\infty} dx_4' x_4'\int_0^\infty dt 
    \frac{K_1(2\sqrt{M^2+t}x_4')}{\sqrt{t}\sqrt{M^2+t}}
    \nonumber\\
    &=& \frac{1}{8M}
  + \frac{M}{\pi}\int_0^\infty dt 
  \frac{1}{\sqrt{t}\sqrt{M^2+t}}
  \frac{\pi}{8(M^2+t)}
  \nonumber\\
  &=& \frac{3}{8M},
\end{eqnarray}
where we have performed the integration in the order of $x_4, x_4'$, and $t$.

\section{Orthogonality of the bulk and edge modes}
\label{app:orth}

In this work, we use different eigensets
in the evaluations of the bulk and edge modes.
Therefore, the orthogonality is generally lost.
In this appendix we show, however, that
the orthogonality is recovered in the continuum limit.

Let us consider the free  bulk fermion mode $\phi_-^\omega(x_4)$
in Eq.~(\ref{eq:bulk}) and
the edge mode $\phi_-^\mathrm{edge}(x_4)$ in Eq.~(\ref{eq:edge'}).
Since the nontrivial link-variable-dependent part is simply neglected,
the edge mode wave function is slightly different from Eq.~(\ref{eq:edge}).
Their inner product is evaluated as
\begin{eqnarray}
  \label{eq:be}
  a\sum_{x_4} \phi^\mathrm{edge}_-(x_4)^\dagger \phi_-^\omega(x_4)
  &=& a \sqrt{\frac{M(2-Ma)}{2\pi}}\sum_{x_4}
  \left(C_\omega e^{-(K-i\omega)x_4}-\mathrm{c.c.} \right)\nonumber\\
  &=& a \sqrt{\frac{M(2-Ma)}{2\pi}}
  \left(C_\omega\frac{1}{1-(1-Ma)e^{i\omega a}}- \mathrm{c.c.} \right),
\end{eqnarray}
where $\mathrm{c.c.}$ denotes the complex conjugate.

Note that $C_\omega$ is proportional to $1-(1+M_+)e^{i\omega a}$
and $1+M_+ = 1-Ma+\Delta a^2$, where $\Delta$ expresses the
contribution from the Wilson term near the physical pole.
Then Eq.~(\ref{eq:be}) becomes
\begin{eqnarray}
&=& a \sqrt{\frac{M(2-Ma)}{2\pi}}\frac{1}{|1-(1+M_+)e^{i\omega a}|}
  \left(\frac{1-(1-Ma+\Delta a^2)e^{i\omega a}}{1-(1-Ma)e^{i\omega a}}-\mathrm{c.c} .\right)
  \nonumber\\
&=&   a \sqrt{\frac{M(2-Ma)}{2\pi}}\frac{1}{|1-(1+M_+)e^{i\omega a}|}
  \left(\frac{-\Delta a^2 e^{i\omega a}}{1-(1-Ma)e^{i\omega a}}- \mathrm{c.c.} \right),
\end{eqnarray}
which vanishes in the $a\to 0$ limit.


\begin{thebibliography}{99}

\bibitem{Luscher:1981zq} 
  M.~Luscher,
  Commun.\ Math.\ Phys.\  {\bf 85}, 39 (1982).
  doi:10.1007/BF02029132

\bibitem{Hasenfratz:1998ri} 
  P.~Hasenfratz, V.~Laliena and F.~Niedermayer,
  Phys.\ Lett.\ B {\bf 427}, 125 (1998)
  doi:10.1016/S0370-2693(98)00315-3
  
  
\bibitem{Atiyah:1968mp} 
  M.~F.~Atiyah and I.~M.~Singer,
  Annals Math.\  {\bf 87}, 484 (1968).
  doi:10.2307/1970715
  

\bibitem{Ginsparg:1981bj} 
  P.~H.~Ginsparg and K.~G.~Wilson,
  Phys.\ Rev.\ D {\bf 25}, 2649 (1982).
  doi:10.1103/PhysRevD.25.2649
  
\bibitem{Neuberger:1997fp} 
  H.~Neuberger,
  Phys.\ Lett.\ B {\bf 417}, 141 (1998).
  doi:10.1016/S0370-2693(97)01368-3

\bibitem{Luscher:1998pqa} 
  M.~Luscher,
  Phys.\ Lett.\ B {\bf 428}, 342 (1998).
  doi:10.1016/S0370-2693(98)00423-7
  
\bibitem{Kikukawa:1998pd} 
  Y.~Kikukawa and A.~Yamada,
  Phys.\ Lett.\ B {\bf 448}, 265 (1999)
  doi:10.1016/S0370-2693(99)00021-0.


  
  
\bibitem{Luscher:1998kn} 
  M.~Luscher,
  Nucl.\ Phys.\ B {\bf 538}, 515 (1999)
  doi:10.1016/S0550-3213(98)00680-4.
  
\bibitem{Fujikawa:1998if} 
  K.~Fujikawa,
  Nucl.\ Phys.\ B {\bf 546}, 480 (1999)
  doi:10.1016/S0550-3213(99)00042-5.

 
\bibitem{Suzuki:1998yz} 
  H.~Suzuki,
  Prog.\ Theor.\ Phys.\  {\bf 102}, 141 (1999)
  doi:10.1143/PTP.102.141.
  
\bibitem{Adams:1998eg} 
  D.~H.~Adams,
  Annals Phys.\  {\bf 296}, 131 (2002)
  doi:10.1006/aphy.2001.6209.


\bibitem{Atiyah:1975jf} 
  M.~F.~Atiyah, V.~K.~Patodi and I.~M.~Singer,
  Math.\ Proc.\ Cambridge Phil.\ Soc.\  {\bf 77}, 43 (1975)
  doi:10.1017/S0305004100049410;
  Math.\ Proc.\ Cambridge Phil.\ Soc.\  {\bf 78}, 405 (1976)
  doi:10.1017/S0305004100051872;
  Math.\ Proc.\ Cambridge Phil.\ Soc.\  {\bf 79}, 71 (1976)
  doi:10.1017/S0305004100052105.
  
\bibitem{Luscher:2006df} 
  M.~Luscher,
  JHEP {\bf 0605}, 042 (2006)
  doi:10.1088/1126-6708/2006/05/042
  [hep-lat/0603029].

\bibitem{Witten:2015aba} 
  E.~Witten,
  Rev.\ Mod.\ Phys.\  {\bf 88}, no. 3, 035001 (2016)
  doi:10.1103/RevModPhys.88.035001, 10.1103/RevModPhys.88.35001.

  
\bibitem{Fukaya:2017tsq} 
  H.~Fukaya, T.~Onogi and S.~Yamaguchi,
  Phys.\ Rev.\ D {\bf 96}, no. 12, 125004 (2017)
  doi:10.1103/PhysRevD.96.125004.

\bibitem{Jackiw:1975fn} 
  R.~Jackiw and C.~Rebbi,
  Phys.\ Rev.\ D {\bf 13}, 3398 (1976).
  
\bibitem{Callan:1984sa} 
  C.~G.~Callan, Jr. and J.~A.~Harvey,
  Nucl.\ Phys.\ B {\bf 250}, 427 (1985).
  
\bibitem{Kaplan:1992bt} 
  D.~B.~Kaplan,
  Phys.\ Lett.\ B {\bf 288}, 342 (1992)
  [hep-lat/9206013].
  
\bibitem{Fukaya:2019qlf} 
  H.~Fukaya, M.~Furuta, S.~Matsuo, T.~Onogi, S.~Yamaguchi and M.~Yamashita,
  arXiv:1910.01987 [math.DG].

  
\bibitem{Itoh:1987iy} 
  S.~Itoh, Y.~Iwasaki and T.~Yoshie,
  Phys.\ Rev.\ D {\bf 36}, 527 (1987).
  doi:10.1103/PhysRevD.36.527

\bibitem{Kaplan:1995pe} 
  D.~B.~Kaplan and M.~Schmaltz,
  Phys.\ Lett.\ B {\bf 368}, 44 (1996)
  doi:10.1016/0370-2693(95)01485-3.

  
\bibitem{Shamir:1993zy} 
  Y.~Shamir,
  Nucl.\ Phys.\ B {\bf 406}, 90 (1993)
  doi:10.1016/0550-3213(93)90162-I.
  
\bibitem{Furman:1994ky} 
  V.~Furman and Y.~Shamir,
  Nucl.\ Phys.\ B {\bf 439}, 54 (1995)
  doi:10.1016/0550-3213(95)00031-M.

\bibitem{Hatsugai:1993ywa} 
  Y.~Hatsugai,
  Phys.\ Rev.\ Lett.\  {\bf 71}, no. 22, 3697 (1993).
  doi:10.1103/PhysRevLett.71.3697;
  Phys.\ Rev.\ B {\bf 48}, no. 16, 11851 (1993).
  doi:10.1103/PhysRevB.48.11851
  


  
\end{thebibliography}
\end{document}